\documentclass[11pt]{amsart}
\usepackage{latexsym,amsfonts}
\usepackage{amssymb,amsmath}
\newcommand{\naturali}{{\hbox{{\rm N}\kern-.2em\hbox{\rm R}}}}
\newcommand{\interi}{{\ \hbox{{\rm Z}\kern-.6em\hbox{\bf Z}}}}
\newcommand{\reali}{{\hbox{{\rm I}\kern-.2em\hbox{\rm R}}}}
\newtheorem{proposition}{Proposition}
\newtheorem{theorem}{Theorem}

\newcommand{\complessi}{{\mathbb{C}}}

\newtheorem{lemma}{Lemma}
\begin{document}
\large


\begin{center}
{\huge{\bf On the Spectrum}}

\vskip 1 cm 
{\huge{\bf of Holonomy Algebras}}

\vskip 2 cm
{\large{\sf   Maria Cristina Abbati and Alessandro Mani\`a}}

\vskip 1 cm
Dipartimento di Fisica,
Universit\`a degli Studi di Milano and

 Istituto Nazionale di Fisica Nucleare,
Sezione di Milano,

via Celoria 16, 20133 Milano, Italy.

E-mail: ``maria.cristina.abbati@mi.infn.it''

``alessandro.mania@mi.infn.it''

\vskip 1 cm  
\end{center}

\noindent {\bf Abstract.}  Connections on a trivial bundle $ M\times G$ can be
 identified with their holonomy maps, i.e. with homomorphisms  of a
groupoid of  paths in {\sl M} into the gauge group $G$.
  For a connected compact $G$, various algebras depending on  the set
 $\mathcal A$ of the smooth connections through their holonomy maps have been
 introduced in the literature, called cylindrical and holonomy algebras. 
  We discuss the relations between these algebras and the consistence of
  their spectra. 
 

\section{Introduction}
 
In the canonical treatment of Yang-Mills theories and, in general,
of gauge theories, the space ${\mathcal A}/Gau (P)$ represents  the space of the  
physical configurations of the system. Here ${\mathcal A}$ denotes the set of
 the smooth connections on a principal bundle $ P(M,G)$ and 
 $Gau (P)$  the group of gauge transformations, the base manifold $M$ is  
connected and $G$ is a connected compact Lie group.

 A moment of paramount importance in developing the canonical quantization program
 for gauge theories  invariant under diffeomorphisms, as proposed by Ashtekar et
 al. (\cite{AshtekarIsham},  \cite{Ashtekar}), is to construct  a
 compactification of the configuration space  ${{\mathcal{A}}/Gau(P)}$. This 
compactification  $\overline{{\mathcal{A}}/Gau(P)}$, called quantum configuration 
space, is achieved by means of the immersion of ${\mathcal A}/Gau (P)$ in the
 spectrum of a suitable $\complessi ^*$-algebra  of functions on
 ${\mathcal A}/Gau(P)$, depending on holonomies on paths or on  loops. 
To obtain the physical states for the corresponding quantum theory,  a measure 
$\mu$ on the spectrum $\overline{{\mathcal A}/Gau(P)}$ is given and  
diffeomorphism invariant states are selected  in the Hilbert  space
 $L^2(\overline{{\mathcal A}/Gau(P)},\mu)$.   This quantization procedure is 
known as the loop quantization and originates from the works of Rovelli and  
Smolin \cite{Rovelli}. 

The starting point is the identification of a connection $A$ with its holonomy map
  $\gamma \mapsto H_A(\gamma )$, where $H_A(\gamma )$ means the parallel transport
 along the path $\gamma$; gauge equivalence classes of connections are identified
 with gauge equivalence classes of holonomy maps on loops (  \cite{Anandan},
 \cite{giles}).

Well known  examples of gauge invariant  functions depending on holonomies are the
 Wilson functions, defined - up to a scalar factor - as the maps
 $A\mapsto  Tr H_A (\lambda )$, where  $\lambda $ is a loop. The generated
 $\complessi ^*$-algebra   is called  holonomy algebra and depends on the 
 differentiability class of loops. In the case that $G$ is $U({\sf n})$ or 
 $SU({\sf n})$, the Wilson functions are separating on ${\mathcal A}/Gau(P)$, so
  that $ {\mathcal A}/Gau(P)$ is densely immersed in the  spectrum of the holonomy
  algebra.

For a general group $G$ the Wilson functions are no more separating, so one
 considers  cylindrical functions, functions of the form 
$A \mapsto f(H_A (\gamma _1 ), ...,H_A (\gamma _n)) $, for given paths
 $\gamma _1,..., \gamma _n$. The $\complessi^*$-algebra generated by the
 cylindrical functions is called cylindrical algebra; the invariant cylindrical
 algebra is also defined. The cylindrical functions  are separating on 
$\mathcal A$ so that ${\mathcal A}$ is densely embedded in the spectrum of the
 cylindrical algebra and, as well,  ${\mathcal A}/Gau(P)$ is densely embedded in
 the  spectrum of the invariant cylindrical algebra.

 The case of piecewise analytic loops and paths was  the first to be  investigated
 and many and nice results were obtained (see   \cite{AshtekarLewandowski}, 
 \cite{ALpro}, \cite{Baez1}, \cite{BaezSpin}, \cite{Misuranulla}). It was proved 
 that the spectrum $\overline{\mathcal A}$ of the cylindrical algebra agrees with
 the space  $Hom({\sf Path}(M),G)$ of the generalized connections, i.e. 
homomorphisms into $G$ of the groupoid ${\sf Path}(M)$ of piecewise 
analytic paths.  The spectrum  $\overline{{\mathcal A}/Gau(P)}$ of the invariant
analytical cylindrical algebra  was proved to agree with the space 
$Hom({\sf Loop}_\star(M),G)/AdG$ of the $AdG$-equivalence classes  of 
 homomorphisms into $G$ of the group  of  piecewise analytic loops.  Measures
 invariant under  analytic diffeomorphisms have been constructed  on this space
 by projective techniques using families of  measures labelled by embedded graphs.
 For the so called natural measure a complete  orthonormal set of  states - the 
spin network states,  depending on  embedded  graphs - was constructed  and the 
invariance under analytic diffeomorphisms implemented.

However the analytic setting is not satisfactory from the physical point of view, 
since  invariance with respect to smooth diffeomorphisms is needed to use this
 quantization scheme for gravity.

In \cite{BaezSawin1} the case of piecewise  smoothly immersed paths was approached
 and   the webs were introduced, special families of paths which  play in
 the smooth immersive setting an analogous role to that of the embedded graphs in
 the  analytic setting.   Like every finite family of piecewise analytic paths
 depends on an embedded graph,  i.e. every path can be written as a composition of
 edges belonging to the graph  or of their inverses, so in the smooth immersive 
setting every finite family of paths depends on a web. This  implies that the 
smooth immersive cylindrical algebra $Cyl(\mathcal A)$ is the limit of the
 algebras  $Cyl_W(\mathcal A)$ generated by the cylindrical functions depending 
on a web $W$;  dually,  the spectrum  of the cylindrical algebra is the  
projective limit of the spectra of $Cyl_W(\mathcal A)$. In  \cite{BaezSawin2} a
 diffeomorphism invariant measure  was constructed using projective techniques,
 spin network states and spin web states were defined. The theory is quite 
involved, due to the fact that webs have a more involved behaviour than graphs
 and that the spectrum of the cylindrical algebra is not so simple to be
 characterized.

 In this paper we introduce the cylindrical algebras, the invariant cylindrical
 algebras and the  Wilson algebras in a general setting and illustrate their
 relations.  These are algebras of functions defined on a subset 
 $\mathcal A$ of $Hom(\Lambda,G) $, the space of homomorphisms of a general
 groupoid $\Lambda$ into a connected compact Lie group $G$. We prove that the
  spectrum of the cylindrical algebras is always the closure 
 $\overline{\mathcal A}$ of  ${\mathcal A}$ in $Hom(\Lambda,G)$.

For applications to gauge theories $\Lambda$ is  a suitable groupoid  of paths.
 The problem arises when $\overline{\mathcal A}=Hom(\Lambda ,G)$, where 
$\mathcal A$ is the set of connections.  In the smooth immersive  case  this is
 true  for a  connected compact semisimple Lie group $G$, as proved in  
\cite{LewThiemann}, but it is not true in the non semisimple case \cite{Fleish}. 
 We  characterize $\overline{\mathcal A}$ as a subset of 
$ Hom(\Lambda ,G)$ for a general  connected compact Lie group $G$, where 
$\Lambda$ is  the groupoid  $Path(M)$ of piecewise smooth immersed paths.
 
  The paper is organized as  follows. In section 2 we introduce the cylindrical 
 algebras  and we formulate in various forms the Approximation Condition, i.e. the
 condition that  $\mathcal A$ is dense in $Hom(\Lambda,G) $. We consider also the 
cylindrical algebras invariant under the action of $G^M$, where $M$ is the space 
of units of $\Lambda$ and we discuss their spectrum. In this section we develop
 and generalize some ideas proposed in \cite{Baumgartel}.

In section 3   we define the Wilson functions  on a subset $\mathcal A$ of 
$Hom(\Lambda,G)$, where $\Lambda$ is a group. We investigate  the spectrum of
 the generated $\complessi^*$-algebra and its relations with the invariant
 cylindrical algebras.

In section 4 we apply the previous results to the holonomy algebras and to
 the cylindrical algebras on the space  ${\mathcal A}$ of  connections on the
 trivial bundle $M\times G$. The groupoid is now  ${\sf Path}(M)$ or  $Path(M)$. 
Also the group ${\sf Loop}_\star(M)$ of piecewise analytic loops or the group  
$Loop_\star(M)$  of piecewise smoothly immersed loops are considered. We discuss 
the relationship between the corresponding  cylindrical algebras and the 
consistence of their spectrum. 

 In the last section we discuss cylindrical algebras in the general setting of non
 trivial bundles.

\vskip 2mm

\section{Algebras of cylindrical functions} 

\vskip 2 mm 
  Let we start  recalling some well known results on Abelian 
$\complessi ^*$-algebras of functions.  Let $X$ be a non empty set, $B(X)$ the
 Abelian $\mathbb{C}^*$-algebra of  bounded functions on $X$ and let 
 ${\mathcal F}\subset B(X)$ a $\complessi ^*$-subalgebra separating on $X$.  
 The evaluation map  associates to every $x\in X$ the multiplicative  functional
 $ev _x:{\mathcal F} \to {\bf C}$, $ev _x (f)= f(x )$. By the assumption that 
${\mathcal F}$ is separating the map $ ev : x \mapsto ev_x$ is an embedding of
 $X$ in $Spec ({\mathcal F})$.  Moreover  $X$ is dense in $Spec({\mathcal F})$ by 
the normality axiom. Therefore  the following proposition holds.

\begin{proposition}\label{embedding} Let  $\mathcal{F}$ be any
 $\mathbb{C}^*$-subalgebra of $B(X)$, separating on
 $X$. Then $X$ is injectively and densely embedded in $Spec ({\mathcal F})$
 by  the evaluation map.
\end{proposition}

\vskip 2 mm
 Let $X$ be a compact (Hausdorff) space. If  $\mathcal F$ is the algebra $C(X)$ of the
continuous functions, the evaluation map is an homeomorphism  of $X$ onto
 $Spec({\mathcal F})$. 

For every nonempty subset $Y$ of $X$, we associate the  ${~}^*$-algebraic
 homomorphism  $R_Y : C(X) \to C(Y)\cap B(Y)$,  $R_Y (F)= F_Y$ where $F_Y$ is the
 restriction  of $F$  to $Y$.  When $Y$ is closed,  $R_Y$ is onto $C(Y)$ by Tietze
 extension Theorem; for $Y\subset Z \subset X$ and $Z$ closed, the range of $R_Y$ 
agrees with the set of restrictions to $Y$ of continuous functions on $Z$. We 
denote by  ${\mathcal F}(Y)$ the  $\complessi ^*$-algebra
 generated by the range of $R_Y$. When $\overline{Y}=X$, the map $R_Y$ is 
an isometric isomorphism of $C(X)$ onto ${\mathcal F}(Y)$ whose inverse  is the
  map which  extends each $F\in {\mathcal F}(Y)$ to a continuous function 
  on $X$. So we have that,  for every $Y\subset X$, the $\complessi ^*$-algebra
 ${\mathcal F}(Y)$ is naturally isomorphic to $C(\overline{Y})$.

\begin{lemma}\label{spettro} For every $Y\subset X$, the map  
$E_Y: Spec({\mathcal F}(Y) ) \to X~, E_Y (\varphi )= x_\varphi$, with
 $x_\varphi$ given by
$$
F( x_\varphi )= \varphi (F_Y) \quad \forall F \in {\mathcal S}\subset  C(X)
$$  
for ${\mathcal S}$ separating on $X$, does not depend on ${\mathcal S}$ and  is a
 continuous embedding  and a homeomorphism onto  $\overline{Y}$.
\end{lemma}

\vskip 1 mm
{\em Proof.} The map $E_Y$ is the composition of $R_Y ^\dagger : 
Spec ({\mathcal F}(Y )) \to Spec (C(X))$ with the inverse of  the evaluation map
 $ev :  X \to Spec (C(X))~.$ The map $R_Y ^\dagger$ is 
injective, since the image  of $R_Y$ is dense in ${\mathcal F}(Y)$ and the 
evaluation map is a homeomorphism, since $X$ is compact. To prove  continuity,
 recall that the  topology on $X$ agrees with the $w^*$-topology. Let 
$\overline{x}= E_Y (\overline{\varphi })$ for a given
 $\overline{\varphi} \in Spec({\mathcal F}(Y))$; consider a finite family 
$\{F_k\}_{k=1,...,r}$ in $C(X)$ and the open neighborhood of $\overline{x}$ 
$$
 \{ x \in X ~|~ |F_k (x)- F_k (\overline{x}) | < \epsilon \quad k=1,...,r \}~. 
$$
Its inverse image by $E_Y$ is the set 
$$
\{ \varphi \in Spec({\mathcal F}(Y)) ~|~
|\varphi ( R_Y F_k )- \overline{\varphi} (R_Y F_k )| <\epsilon 
 \quad k=1,...,r\}~,$$
 an open set in the $w^*$-topology. So $E_Y$ is continuous, its image is closed
 and contains $\overline{Y}$, since $E_Y (ev_y)= y$ for every $y \in Y$.

 By 
Proposition \ref{embedding}, every 
$\varphi \in Spec ({\mathcal F}(Y))$ is the weak limit of some net of pure states
 $\{ \varphi _\mu \}$ with $\varphi _\mu = ev_{y_\mu}$, $y_\mu \in Y$, hence  
 $E_Y (\varphi )= \lim_\mu y_\mu$, proving that $E_Y$ is onto $\overline{Y}$.

 Finally, we recall that every continuous injection of a compact space into a
 Hausdorff topological space is a homeomorphism with its image.
\hspace{\stretch{1}}$\Box$

 \vskip 2 mm 
The condition   that $Y$ is dense in $X$ will be called  {\em Approximation
 Condition on $Y$} and  is conveniently stated in the following way:
\par 
\vskip 1 mm
{\em For every  $ x\in X$, every finite family 
 $\{ F _k \}_{k=1,...,r}$ in $C(X)$ and  $\epsilon  >0$
 there exists $y \in Y$ such that 
$$
 | F_k (x) - F_k (y)| <\epsilon  \quad k=1,...,r~.
$$
}
 The  condition can be restated by choosing $\{ F_k \}_{k=1,...,r} \subset
 {\mathcal S}$ if  ${\mathcal S}$ is any separating subset of $ C(X)$. 

 \vskip 2 mm
  Now we assume that a compact group ${\sf G}$ acts continuously on $X$. The
 quotient space $X/{\sf G}$ is a (Hausdorff) compact space and the canonical 
projection $[~]~: X \to X/{\sf G}$ is continuous, open and closed.  The
 $\complessi^*$-algebra $C_{\sf G}(X)$ of the ${\sf G}$-invariant continuous
 functions on $X$ is identified with $C(X/{\sf G})$ by pull-back with the
 projection. As $X/{\sf G}$ is compact,  $C_{\sf G} (X)$ is separating on 
 $X/{\sf G}$. 

 For every ${\sf G}$-invariant subset $Y$ of $X$ we get by simple topological
 arguments  that $\overline{Y}$ is ${\sf G}$-invariant, that
 $\overline{Y/{\sf G}}= \overline{Y}/{\sf G}$ and that  the map $R_Y$ is
 equivariant.  We denote by ${\mathcal F}_{\sf G} (Y)$  the
 $\complessi ^*$-subalgebra of  ${\sf G}$-invariant functions in
 ${\mathcal F}(Y)$. 

\begin{lemma}\label{invar} For every ${\sf G}$-invariant $Y\subset X$,
 the map  $ I_Y: Spec({\mathcal F}_{\sf G}(Y) ) \to X/{\sf G}$, 
$I_Y (\varphi )= [x_\varphi ]$, where $x_\varphi \in X$ satisfies
$$
F( x_\varphi )= \varphi (R_Y F) \quad \forall F\in{\mathcal S} \subset
 C _{\sf G} (X)
$$
 for  ${\mathcal S}$ separating on $X/G$,  does not depend on ${\mathcal S}$ and
  is a continuous embedding and a homeomorphism onto  $\overline{Y/{\sf G}}$.
  $I_Y$ is a homeomorphism onto $X/{\sf G}$ if and only if $Y$ satisfies the 
 Approximation Condition.
\end{lemma}

{\em Proof.} We identify $C_{\sf G}( X)$ with $C(X/{\sf G})$ and 
 ${\mathcal F}_{\sf G} (Y)$ with ${\mathcal F}(Y/{\sf G})$; then we can apply
  Lemma \ref{spettro} to ${\mathcal F}(Y/G)$. 

 For the last statement, we remark that 
$\overline{Y /{\sf G}}= \overline{Y}/{\sf G}$ and equals $X/{\sf G}$ if
 and only if $\overline{Y}=X$. This   follows easily by the ${\sf G}$-invariance of
 $\overline{Y}$.\hspace{\stretch{1}} $\quad \Box$

\vskip 2 mm
As an immediate consequence of the previous arguments we obtain the canonical
 isomorphisms
$$
{\mathcal F}_{\sf G}(Y) \equiv C_{\sf G}( \overline{Y}) \equiv
 C(\overline{Y}/{\sf G})~.
$$

 When $Y$ is ${\sf G}$-invariant the Approximation Condition on $Y/{\sf G}$
 can  be stated on $Y/{\sf G}$ in terms of ${\sf G}$-invariant functions, as
 follows: 
\par 
\vskip 1 mm
{\em For every  $ x\in X$, every finite family 
 $\{ F_k \}_{k=1,...,r} \subset C _{\sf G} (X)$ and  $\epsilon  >0$
 there exists $y \in Y$ such that
$$
 |F _k (x) ) - F_k (y)| <\epsilon  \quad k=1,...,r~.
$$
}
 The condition can be restated by choosing the functions $F_k$ in any separating 
 subset ${\mathcal S} \subset C_{\sf G}(X)$.

More generally, let $\overline{Y}$  be ${\sf G}$-invariant but $Y$ be not
 necessarily ${\sf G}$-invariant.  This can be true, e.g., if the Approximation 
Condition on $Y$ holds. Using the isomorphism 
${\mathcal  F}(Y)\equiv C(\overline{Y})$, we can
 again consider the subalgebra ${\mathcal F}_{\sf G} (Y)$ of the functions 
$F \in {\mathcal F}(Y)$ whose extension is ${\sf G}$-invariant and  we have
$$
 Spec ({\mathcal F}_{\sf G} (Y))\equiv \overline{Y}/{\sf G}~.
$$

 \vskip 3mm

 Now we come to the cylindrical algebras. Let us begin with the definition 
 of groupoid. 

 A groupoid is a set $\Lambda$ endowed with a binary composition
 law  satisfying:

i) to every $\lambda \in \Lambda$ an element   
 $\lambda ^{-1}$ (the inverse) is associated such that $r(\lambda )=
\lambda \lambda ^{-1}$ and $s(\lambda )= \lambda ^{-1} \lambda $ exist and are
 the right and the left unit of $\lambda$, respectively; 

ii) for $\lambda ,\eta \in \Lambda$ the product $\lambda \eta$ exists if and 
only if $r(\eta ) = s(\lambda)$;

iii) when defined, the product is associative.

\vskip 2 mm 
 We  denote by $M(\Lambda )$ the set of units of $\Lambda$, i.e. the elements
 of $\Lambda$ of the form $\lambda \lambda ^{-1}$ for some $\lambda \in \Lambda$.
 A groupoid  $\Lambda$ is a group if and only if  $M(\Lambda )$ is a singleton. 
\vskip 2mm
Let $G$ denote  a closed subgroup of $U({\sf  n})$, the group of unitary matrices 
in dimension ${\sf n}$.  The set  of homomorphisms $ H: \Lambda \to G$,  denoted
 by  $Hom (\Lambda , G)$, is a compact space since it is  closed in   $G^\Lambda$,
 a compact  group in  the Tychonoff product topology. 

\vskip 2 mm

For a  continuous function $f: G^m \to {\complessi}$ and  a finite family
 $\{\lambda _k \}_{ k=1,...,m}$ in  $\Lambda $,   the continuous function
 $F_{\lambda _1 ,...,\lambda _m  ;f} : Hom (\Lambda , G) \to {\complessi}$
 defined  by
$$
 F_{\lambda _1, ...,\lambda _m ;f}(H) = f(H(\lambda _1 ),H(\lambda _2 ),...,
H(\lambda _m ))~,
$$
 is called {\em cylindrical function}. The  cylindrical
 functions form  a normed $*$-algebra $cyl(Hom(\Lambda,G))$, whose completion is 
 a  $\complessi ^*$-algebra, denoted by  $Cyl (Hom (\Lambda , G))$. For any  subset
 ${\mathcal A}$  of $Hom( \Lambda , G)$, the *-algebra  of the restrictions to 
${\mathcal A}$ of the cylindrical functions is denoted by  $cyl ({\mathcal A})$ 
and the  completion of $cyl({\mathcal A})$ by $Cyl ({\mathcal A})$.

\vskip 2 mm
 For $i,j=1,...,{\sf n}$ and $\lambda \in \Lambda$, 
the cylindrical function $\Phi _{i,j;\lambda }$
 defined  by
$$
 \Phi _{i,j;\lambda } (H)= H( \lambda )_{i,j}~\quad H\in Hom (\Lambda , G)~,
$$
where $H( \lambda )_{i,j}$ denotes the corresponding matrix elements of
 $ H( \lambda )$,  will be called representative function. The representative
 functions are separating on $Hom (\Lambda , G)$, hence
 their restrictions  are separating on every subset 
${\mathcal A}$.  By Proposition \nolinebreak[4]\ref{embedding}
   the evaluation  map is a dense 
embedding  of ${\mathcal A}$ in $Spec (Cyl ({\mathcal A}))$.

When ${\mathcal A}$ is closed,   the Weierstrass Theorem gives  
$Cyl({\mathcal A})= C({\mathcal A})$; as a special case we have 
$Cyl (Hom (\Lambda , G))= C(Hom (\Lambda , G))$. Recalling that 
 ${\mathcal F}({\mathcal A})$ is the $\complessi^*$-algebra  
 generated by the range of the   restriction map 
$R_{\mathcal A} :C(Hom (\Lambda , G)) \to C({\mathcal A})$, we get easily that
  $ {\mathcal F}({\mathcal A})=Cyl ({\mathcal A})$. As a consequence of
 Lemma \ref{spettro} we obtain the following theorem. 

\begin{theorem}\label{cilindriche} For every
 ${\mathcal A} \subset Hom (\Lambda , G)$,  the map 
 $E_{\mathcal A} : Spec(Cyl({\mathcal A})) \to Hom (\Lambda , G)$, 
 $E_{\mathcal A} (\varphi )= H_\varphi$, 
 with $H_\varphi : \Lambda \to G$ given by 
$$
 \Phi_{i,j;\lambda } (H_\varphi )= \varphi (\Phi _{i,j;\lambda})
$$
 for every representative function $\Phi_{i,j;\lambda }$, is a continuous embedding
 and a  homeomorphism onto  $\overline{\mathcal A}$. The map $E_{\mathcal A}$
 is  onto $Hom(\Lambda , G)$ if and only if  the Approximation
 Condition on ${\mathcal A}$ is satisfied.
\end{theorem}

 In this setting, the Approximation Condition on  $\mathcal A$ is conveniently
 stated in the following way:
\par 
\vskip 1 mm
{\em For every  $ H$ in $ Hom (\Lambda , G)$, every finite family 
 $\{ \lambda _k \}_{k=1,...,r}\subset \Lambda$ and  $\epsilon  >0$, 
 there exists $H_A \in {\mathcal A}$ such that
$$
 \|H(\lambda _k ) - H_A (\lambda _k )\| <\epsilon  \quad k=1,...,r~.
$$
}
 The condition can be also restated as follows:

\vskip 2 mm 
{\em For every  $ H$ in $ Hom (\Lambda , G)$, every finite system 
of representative functions $\{\Phi _{i_k, j_k ;\lambda _k}\}_{k=1,...,r}$ and 
 $\epsilon  >0$, there exists $H_A \in {\mathcal A}$ such that

\vskip 1 mm
$$
 |\Phi _{i_k ,j_k ;\lambda _k}(H) - \Phi _{i_k, j_k ; \lambda _k}(H_A )|
<\epsilon  \quad k=1,...,r~.
$$
}
\par
\vskip 1 mm 
The Tychonoff and the $w^*$-topology agree  on $Hom(\Lambda,G)$. In the first
 version of the Approximation Condition, the density
 of ${\mathcal A}$ in $Hom(\Lambda,G)$ is expressed  in terms of the usual basis
 for the Tychonoff topology, in the second one using   a basis  for the 
$w^*$-topology.

\par
\vskip 3 mm
  From now on, we  denote $M(\Lambda )$ shortly by $M$ and consider the natural
 continuous right action of the compact group  $G^M $ on $Hom (\Lambda ,G)$ given
 by 
$$
  (H.g)(\lambda ) =g^{-1}(r(\lambda ))H(\lambda ) g(s(\lambda ))~.
$$
 
 The dual isometric action on $C(Hom (\Lambda  ,G))$ is defined by
 $gF (H)= F(H.g^{-1})$ for every $H \in Hom(\Lambda , G)$ and $g \in G^M$. 
 For a $G^M$-invariant    ${\mathcal A} \subset Hom (\Lambda , G)$, we denote by  
 $Cyl _{G^M} ({\mathcal A})$  the $\complessi ^*$-subalgebra of the 
${G}^M$-invariant functions of $Cyl({\mathcal A})$.

By Lemma \ref{invar} we obtain the following theorem.
 
\begin{theorem}\label{stati1} Let   ${\mathcal A}$ be
 a ${G}^M$-invariant subset of $Hom (\Lambda , G)$. The map  
$ I_{\mathcal A}: Spec(Cyl_{G^M}({\mathcal A}) ) \to Hom (\Lambda , G)/G^M$, 
$I_{\mathcal A} (\varphi )=[H_\varphi ]$, where $H_\varphi$ satisfies
$$
F( H_\varphi )= \varphi (R_{\mathcal A}F) \quad 
\forall F\in {\mathcal S} \subset Cyl _{G^M} (Hom (\Lambda ,G))
$$
for  $\mathcal{S}$ separating on $(Hom (\Lambda ,G)/G^M$, does not depend on
 ${\mathcal S}$ and   is a continuous 
embedding and a homeomorphism onto $\overline{\mathcal A}/G^M$. The map
 $I_{\mathcal A}$ is onto $Hom (\Lambda , G)/{ G^M}$ if and only if the 
 Approximation Condition on ${\mathcal A}$ is satisfied.
 \end{theorem}

 \vskip 2 mm 
In the special case that $\Lambda$ is a group, the set of units is  just $\{ e\}$,
 so $G^M=G$ and its action on $Hom (\Lambda , G)$ is given by 
$(H.a)(\lambda )= a^{-1}H(\lambda ) a= Ad _{a^{-1}} H(\lambda )$ for $a \in G$. For
 any $Ad G$-invariant ${\mathcal  A}$, we have 
$$
 Spec (Cyl_{Ad G} ({\mathcal A})) =  Spec(Cyl ({\mathcal A})) /Ad G = 
\overline{\mathcal A}/Ad G~.
$$

\vskip 2 mm
  A natural choice for ${\mathcal S}$ in Theorem \ref{stati1} is 
 $cyl _{G^M} (Hom (\Lambda , G))$, the  *-sub\-al\-ge\-bra of the 
 $G^M$-invariant cylindrical functions. Let us show that
 $cyl _{G^M} (Hom (\Lambda ,G))$ is separating on $Hom (\Lambda ,G)/G^M$.

\begin{theorem}\label{invarianti} Let $dg$ denote the normalized Haar measure
 on $G^M$. 
\par
i) The mean value map
$
\langle ~\rangle : C(Hom (\Lambda , G)) \to C _{G^M} (Hom (\Lambda , G))
$
 defined by
$$
 \langle F \rangle = \int _{G^M} gF dg
$$
 restricts to a continuous surjection from  $cyl (Hom (\Lambda , G))$ onto
 the $*$-subalgebra $cyl _{G^M} (Hom (\Lambda , G))$ of $G^M$-invariant
 cylindrical functions.

\par
ii) The $*$-algebra $cyl _{G^M} (Hom (\Lambda , G))$ is separating on
 $Hom (\Lambda , G) / G^M$.
\end{theorem}

{\em Proof.} i) The only non trivial point in the first statement is that
 the mean value map sends  $cyl (Hom (\Lambda , G))$ into 
$cyl _{G^M} (Hom (\Lambda , G))$. Actually, for every cylindrical function
$F=F_{\lambda _1 ,...,\lambda _m ;f}$ we have  
$$
(gF)(H)= f(g(r(\lambda _1 ))H(\lambda _1)g^{-1}(s(\lambda _1)),..., 
g(r(\lambda _m ))H(\lambda _m)g^{-1}(s(\lambda _m)))~.
$$
 The elements of $G^{2m}$ of the form 
 $(g(r(\lambda _1)), g(s(\lambda _1 )),...., g(r(\lambda _m)), g(s(\lambda _m )))$
 for some $g \in G^M$ form a subgroup $\Gamma$ isomorphic to $G^d$, where
 $d$ denotes the number of distinct units in $M$ which are
  of the form   $r(\lambda _i )$ or $s(\lambda _i )$ for some $i=1,..,m$. Therefore
\begin{displaymath}
\langle F \rangle (H) = \int _{G^M} f(g(r(\lambda _1 ))H(\lambda _1)g^{-1}
(s(\lambda _1)),..., g(r(\lambda _m ))H(\lambda _m)g^{-1}(s(\lambda _m)))  dg
\end{displaymath}
\begin{displaymath}
 =\int _{\Gamma }(a^{-1} f)( H(\lambda _1), ....,H(\lambda _m)) da
 \end{displaymath}
 where for $a\in G^{2m}$, $a^{-1}f$ is given by  $a^{-1}f(g_1,...,g_{m})= 
f(a_1 g_1 a_2 ^{-1}, ..., a_{2m-1}g_m a_{2m})$ 
 and  $da$ denotes the Haar measure on $\Gamma \equiv G^{d}$.
  Thus $\langle F\rangle$  is a cylindrical function, since 
$ \langle F \rangle = F _{\lambda _1 ,...,\lambda _m ;\tilde{f}}$
 where $\tilde{f}$, given by 
$ \tilde{f} (g_1,...,g_m)= \int _{\Gamma }(a^{-1} f)( g _1, ...., g _m ) da$, 
 is continuous.

 ii) Let $F_n \in cyl (Hom (\Lambda , G))$ converge to 
 $F\in Cyl _{G^M} (Hom (\Lambda ,G))$. Then 
  $\langle F_n \rangle  \to\langle F\rangle =F$, proving that  
$cyl_{G^M} (Hom (\Lambda , G))$ is dense in  $Cyl _{G^M}(Hom (\Lambda , G))$.
 This is equivalent to say that $cyl_{G^M} (Hom (\Lambda , G))$ is separating.
\hspace{\stretch{1}}$\Box$

\vskip 3 mm

 Fix a point $\star$ in $M$. The elements $\lambda$ of $\Lambda$ 
satisfying
 $s(\lambda )= r(\lambda )= \star$  form a group, denoted by $\Lambda _\star$. 
 The restriction map  defines  a continuous projection 
$$\mathcal{P}_\star : Hom(\Lambda ,G) \to Hom(\Lambda _\star , G)~,$$ 
which  is   equivariant w.r.t. the action of $G^M$;
 actually $(Hg)_\star=Ad_{g(\star)^{-1}}H_\star$, where
 $H_\star={\mathcal P}_\star(H)$. 

We say that $M$ is $\Lambda$-connected if , for every  $s,r \in M$, there exists
 $\lambda \in \Lambda$ with $s=s(\lambda )$ and $r= r(\lambda )$.  If $M$ is
 $\Lambda$-connected,  ${\mathcal P}_\star$ is onto and quotients to a
 homeomorphism 
$$
{\mathcal Q}_\star:Hom(\Lambda ,G)/G^M \to Hom (\Lambda _\star ,G)/Ad G~.
$$ 

 This is proved following an argument of Velhinho in \cite{Velhinho}. Let
 $G^M _\star$ the compact subgroup of maps $g\in G^M$ such that $g(\star )=1$.
 For every $x\in M$, we fix a unique $e_x \in \Lambda$ with $s(e_x)  =\star$ and 
$r(e_x)=x$, choosing $e_\star =\star$.  To every $H$ we  associate 
 $g_H\in G_\star ^M$ by defining $g_H (x)= H(e_x)$. The map 
$$
\Theta  : Hom(\Lambda , G) \to Hom (\Lambda _\star , G) \times G^M _\star,
 \quad H\mapsto ( H_\star, g_H )
$$
 is a homeomorphism. Actually, ${\mathcal P}_\star$ is onto since for 
$h\in Hom(\Lambda_\star ,G)$ we define $H\in Hom(\Lambda ,G)$ such that
 $H_\star =h$ by $H(\lambda )= h(e_{r(\lambda )} ^{-1} \lambda e_{s(\lambda)})$. 
  The map  $\Theta$  is continuous and it is onto 
since, for every $H \in Hom (\Lambda, G)$ and
$g\in G_\star ^M$, we have
 $\Theta (H.(g_H g^{-1}))= (H_\star, g)$. The inverse is defined by 
$(H_\star,g)\mapsto H^\prime .(g_{H^\prime} g^{-1})$ for any $H^\prime$ such
 that ${\mathcal P}_\star(H^\prime )=H_\star$.

A continuous  action of $G^M$ on  $Hom (\Lambda_\star,G )\times G_\star ^M$ is
 given by
$$
(H_\star,g^\prime ).g=(Ad _{g^{-1}(\star )} H_\star, R_g(g^\prime))~,
$$
where $R_g(g^\prime)(x)=g(x)^{-1}g^\prime (x) g(\star)$. It is clear that 
  $\Theta $ quotients to the wanted homeomorphism ${\mathcal Q}_\star$.

\vskip 3mm

If $\mathcal A$ is $G^M$-invariant, then ${\mathcal A}_\star =
 {\mathcal P}_\star ({\mathcal A})$ is $Ad G$-invariant; the restriction of 
${\mathcal Q}_\star$ is  a homeomorphism of ${\mathcal A}/G^M$ onto 
${\mathcal A}_\star / Ad G$ and, obviously, of $\overline{{\mathcal A}/G^M }$
onto $\overline{{\mathcal A}_\star /Ad G}$. By duality we obtain the isomorphism 
$$
Cyl_{G^M}({\mathcal A}) \equiv   
Cyl _{Ad G} ({\mathcal A}_\star )~.
$$

 If the Approximation Condition on  ${\mathcal A}$ is satisfied,  we get 
$$
 Spec (Cyl_{ G^M} ({\mathcal A})) \equiv Hom (\Lambda _\star , G) /Ad G~.
$$
\vskip 3mm
 We take now into account the case that  $\overline{\mathcal A}$ is 
$G^M$-invariant but ${\mathcal A}$ is not necessarily $G^M$-invariant.

\begin{theorem}\label{vattelapesca}
 If $\overline{\mathcal A}$ is $G^M$-invariant and ${\mathcal A}_\star$
is $AdG$-invariant, then 
$$
Spec(Cyl _{G^M} ({\mathcal A}))\equiv \overline{\mathcal A}/G^M
 \equiv \overline{{\mathcal A}_\star / Ad G} \equiv Spec (Cyl _{Ad G}
 ({\mathcal A}_\star ))
$$
 where $Cyl_{G^M}({\mathcal A})$ denotes the $\complessi ^*$-algebra of 
restrictions to ${\mathcal A}$ of functions in 
$Cyl _{G^M}(\overline{\mathcal A})$.  Moreover
 $\overline{\mathcal A}_\star = Hom (\Lambda _\star ,G)$ if and only if 
$\overline{\mathcal A}= Hom (\Lambda , G)$.
 \end{theorem}

{\em Proof}. By definition, $Cyl _{G^M} ({\mathcal A})$ is isomorphic to
 $Cyl _{G^M} (\overline{\mathcal A})$ whose spectrum is 
$\overline{\mathcal A} /G^M$, since $\overline{\mathcal A}$ is
 $G^M$-invariant. The restriction of  ${\mathcal Q}_\star$
 to
 $\overline{\mathcal A}/G^M$ is a homeomorphism onto
 $(\overline{\mathcal A})_\star /Ad G$.
By continuity of $\mathcal P_\star$, the set
 $(\overline{\mathcal A})_\star$ is closed, hence
 the relation ${\mathcal A}_\star \subset (\overline{\mathcal A})_\star 
\subset \overline{{\mathcal A}_\star }$ implies that 
$(\overline{\mathcal A})_\star = 
\overline{{\mathcal A}_\star}$. 
Moreover the $AdG$-invariance of ${\mathcal A}_\star$
gives $\overline{{\mathcal A}_\star} /Ad G =
 \overline{{\mathcal  A}_\star /Ad G}$.  Hence we have  ${\mathcal Q}_\star(
\overline{{\mathcal A}}/G^M )=\overline{{\mathcal A}_\star/AdG} $.

 To prove  the second statement, let 
$\overline{{\mathcal A}_\star}=Hom(\Lambda_\star,G)$.
This is equivalent to 
   $\overline{\mathcal A}_\star /Ad G =(\overline{\mathcal A})_\star/AdG= 
Hom (\Lambda _\star ,G)/Ad G$.  By the homeomorphism ${\mathcal Q}_\star$ we
 obtain  $\overline{\mathcal A}/G^M = Hom  (\Lambda   ,G) / G^M$.
  The $G^M$-invariance of $\overline{\mathcal A}$ gives 
 $\overline{\mathcal A} = Hom  (\Lambda   ,G)$. \hspace{\stretch{1}}$\Box$

\par
\vskip 2 mm

\section{The Wilson Algebras}

In this section  $\Lambda$ will be a group and $G$ a closed subgroup of 
$U({\sf n})$.  To every $\lambda \in \Lambda$ we associate the  cylindrical map
 $T_{\lambda }$ on $ Hom (\Lambda , G)$ defined by
$$
T_{\lambda } (H)= \frac{1}{{\sf n}}Tr (H(\lambda ))~.
$$
\vskip 1 mm
These  functions are called Wilson functions.
  To every ${\mathcal A} \subset Hom(\Lambda , G)$ we associate
  the {\em Wilson $\complessi ^*$-algebra of ${\mathcal A}$},  
denoted by $\mathfrak{H} _{\mathcal A}$, which is the  $\mathbb{C}^*$-algebra 
generated by  the Wilson functions restricted to ${\mathcal A}$.
 
 If $H,H^\prime\in Hom(\Lambda,G)$ are
 equivalent representations of $\Lambda$ we have 
$T_{\lambda }(H) =T_{\lambda}(H^\prime)$ for every $\lambda\in \Lambda$.  Besides, 
it is well known that  equivalent homomorphisms of any group 
 in $U({\sf n})$ are unitarily  equivalent. So we  consider the quotient of 
$Hom(\Lambda,G)$ by unitary equivalence. For $H \in Hom (\Lambda , G)$ we denote
 by $\widehat H$ its unitary equivalence class and
  by $\widehat{\mathcal A}$ the set of unitary equivalence classes of
 homomorphisms in  ${\mathcal A}$.

 The Wilson functions are separating on $\widehat{Hom}(\Lambda , G)$, as follows
 from the next proposition.

\begin{proposition} \label{nome} Let $\Lambda $ be a  group,
 $H$ and $H^\prime$ in $Hom (\Lambda , G)$. If
 $T_\lambda (H) = T_\lambda (H^\prime )$ for  every  $\lambda \in \Lambda$, 
 then $H$ and $H^\prime $ are unitarily equivalent.
\end{proposition}

 {\em Proof}. To begin with, let us  consider a topological group  $\Lambda$ and
 continuous homomorphisms. We recall that there exists a compact (Hausdorff) group
 $Cpt(\Lambda )$, called  the associated compact  group of
 $\Lambda$, and  a homomorphism $\varkappa : \Lambda \to Cpt(\Lambda )$ 
with dense range  such that the following universality property holds: to  
  any $G$-valued  continuous  representation $H$ of $\Lambda$ one can associate 
 a unique representation $K : Cpt (\Lambda ) \to G$, such that 
 $H =K \circ \varkappa$ (see e.g. \cite{Dixmier}). By density of
 $\varkappa (\Lambda )$ in $Cpt (\Lambda )$,  the equality 
 $T_\lambda (H)  =T_\lambda (H^\prime )$ for every $\lambda \in \Lambda$ 
 implies that $Tr (K(\xi )) =Tr(K^\prime (\xi) )$, for every 
$\xi \in Cpt( \Lambda)$.  By  a well known theorem on the representations of
 compact groups (see e.g.  \cite{Broecker}) $K$ and $K^\prime$ are equivalent. This
 implies that $H$ and $H^\prime$ are  equivalent, hence unitarily equivalent.

  If no topology is assumed  on $\Lambda$, we give $\Lambda$ the 
topology induced by  all  homomorphisms $H: \Lambda\to G$, so that $\Lambda$
 becomes a topological group and $H$ and $H^\prime$  continuous representations. 
Therefore we are reduced to the previous case. \hspace{\stretch{1}}$\quad \Box$

\par
\vskip 2 mm

\vskip 2 mm
 The Proposition \ref{nome} assures that $\mathfrak{H}_{\mathcal A}$ is
  separating on  $\widehat{\mathcal A}$, hence $\widehat{\mathcal A}$
 is densely embedded in $Spec({\mathfrak H} _{\mathcal A})$.
 To characterize  $Spec({\mathfrak H} _{\mathcal A})$ it is convenient to
 consider $\mathcal A$ as a subset of $Hom (\Lambda , U({\sf n}))$ and identify
 $\widehat{\mathcal A}$ with the subset of 
$Hom (\Lambda , U({\sf n}))/Ad U({\sf n})$ obtained by applying to $\mathcal A$  
 the projection $\hat{~}: Hom (\Lambda , U({\sf n}))\to 
Hom (\Lambda , U({\sf n}))/Ad U({\sf n})$. The Wilson functions are separating on 
$Hom (\Lambda , U({\sf n})) /Ad U({\sf n})$, so they generate the algebra of
 continuous functions on $Hom (\Lambda , U({\sf n}))/Ad U({\sf n})$.
Then, by Lemma \ref{spettro}, $Spec({\mathfrak H} _{\mathcal A})$ is homeomorphic
 to the closure of  $\widehat{\mathcal A}$ in 
$Hom (\Lambda , U({\sf n}))/Ad U({\sf n})$.

\vskip 2 mm
  Wilson functions become relevant in gauge theories  when ${\mathcal A}$ is
 $Ad G$-invariant and  $\widehat {\mathcal A}= {\mathcal A}/Ad G $.
To obtain this identification we have to specialize the group $G$. 
 One can  assume that for every  $a\in G$ there exist $\lambda \in \Lambda$ and
 $H\in {\mathcal A}$ such that $a=H(\lambda)$. This is always satisfied in  the
  cases arising in  gauge theories. Under this assumption the condition
$ \widehat{\mathcal A}= {\mathcal A}/Ad G$ implies  that $G$ is a normal subgroup
 of $U({\sf n})$.

  Conversely, if a  group $G$ is a closed normal subgroup of $U({\sf n})$,
 one gets  $\widehat {Hom}(\Lambda , G)= Hom(\Lambda , G)/Ad G$. This is a
  consequence of the following proposition.

\begin{proposition} \label{lemma} Let $G$ a closed normal subgroup $U({\sf n})$. 
 Then the conjugation classes w.r.t.  $Ad_{U({\sf n})}$ and $Ad_G$ coincide on $G$.
 \end{proposition}

{\em Proof.} The smooth group homomorphism $\Phi : U(1) \times SU({\sf n}) \to
 U({\sf n}),\quad \Phi(\alpha , S)= \alpha S $, is onto since, for
  $U\in U({\sf n})$,
 $\Phi (\delta  ,\delta ^{-1}U)=U$ for every  
$\delta \in U(1)$ such that  $\delta ^{\sf n}  = det (U)$.
  As $G$ is a closed normal subgroup of $U({\sf n})$, then
  $\widetilde{G}=\Phi ^{-1} (G)$ is a compact normal subgroup of
 $U(1) \times SU({\sf n})$.  Therefore  $\pi _s (\widetilde{G})$ is a compact
 normal subgroup of $SU({\sf n})$, where 
  $\pi _s : U(1) \times SU({\sf n}) \to SU({\sf n})$ is the canonical projection. 
  As $SU({\sf n})$ is simple, compact and connected,  we have two cases:
 
i) $\pi _s (\widetilde{G})= SU({\sf n})$;

ii) $\pi _s (\widetilde{G})$ is a (finite) central subgroup of $SU({\sf n})$.  

We discuss separately these cases. 

i) Let $U\in U({\sf n})$ of the form $U= \alpha S$ with $\alpha \in U(1)$ and 
 $S\in SU({\sf n})$.  By assumption i), there exists
  $(\beta , S)\in \widetilde{G}$, for some $\beta \in U(1)$, so
 that $h= \beta S$ belongs to $G$. For every $V\in U({\sf n})$ we have
$ Ad_U V= Ad _S V=Ad _h V~.$ 

ii) If $\pi _s (\widetilde{G})$ is a central subgroup of $SU({\sf n})$, then 
$\widetilde{G}$ is a central subgroup of $U(1) \times SU({\sf n})$, so that $G$
 is a central subgroup of $U({\sf n})$. Thus the adjoint actions on $G$ are
 both trivial.    \hspace{\stretch{1}}$\quad \Box$
 
\vskip 2 mm

Then, let  $G$ be a closed normal subgroup of $U({\sf n})$. The Wilson
 functions  become a   separating set of cylindrical functions on  
 $Hom (\Lambda , G)/Ad G=\widehat{Hom}(\Lambda,G)$, 
so they generate the algebra 
$ C_{Ad G} (Hom (\Lambda , G))$. 
For every
 $Ad G$-invariant  subset ${\mathcal A}$ of $ Hom(\Lambda,G)$, we have 
$\widehat{\mathcal A}={\mathcal A}/AdG$ and
   $\mathfrak{H}_{\mathcal A} =Cyl _{Ad G}({\mathcal A})$.

The density of $\widehat{\mathcal A}$ in $\widehat{Hom} (\Lambda , G)$ is assured
 by  the {\em Wilson Approximation Condition on ${\mathcal A}$}, formulated as 
follows:

\vskip 1 mm
{\em For every $ H$ in $ Hom (\Lambda , G)$, every finite family 
 $\{ \lambda _k\}_{k=1,...,r}$ of $ \Lambda$  and
$\epsilon >0$,  there exists $H_A \in {\mathcal A}$ such that

$$
 | Tr  H (\lambda _k) - Tr H_A (\lambda _k )| < \epsilon \quad k=1,...,r~.
$$
}
 Thus we have obtained the following theorem.

\begin{theorem}\label{stati} Let $\Lambda$ be a group, $G$ a closed normal subgroup
 of $U({\sf n})$ and ${\mathcal A}$ an $Ad G$-invariant subset of 
 $Hom (\Lambda , G)$. Then:

i) ${\mathfrak H}_{\mathcal A}$ agrees with $Cyl_{AdG}(\mathcal A)$;

ii)  the map  
$ I_{\mathcal A}: Spec(\mathfrak{H}_{\mathcal A})  \to
 \widehat{Hom} (\Lambda , G)$, $ I_{\mathcal A} (\varphi )= [H_\varphi ]$, 
 where $H_\varphi$ satisfies
$$
\varphi (T_\lambda ) =  
\frac{1}{{\sf n}}Tr H_\varphi (\lambda ) \quad  \lambda \in \Lambda~,
$$
  is a homeomorphism onto the closure of ${\widehat{\mathcal A}}$ in 
 $\widehat{Hom} (\Lambda , G)$;

iii) the embedding $I_{\mathcal A}$ is onto
 $\widehat{Hom} (\Lambda , G)$ if and only if  the Wilson Approximation Condition is satisfied.
\end{theorem}
\vskip 2 mm

\section{Application to Gauge Theories}
\par
\vskip 2 mm

 Now we discuss the applications to gauge theories of the statements proved in the
 above sections. 
 
Let  $M$ be a connected, orientable paracompact smooth manifold with $dim(M)>1$.
 Then $M$ admits a compatible real analytic structure, which is unique up to 
 diffeomorphisms. 

We will start considering  continuous piecewise smooth (or piecewise analytic)
 parametrized paths and  loops  $\gamma : [0,1] \to M$. Paths  $\gamma$ and
 $\lambda$ with $\gamma (1)=\lambda (0)$ can 
 be composed to get $\lambda \gamma : [0,1] \to M$ defined by
$$
(\lambda \gamma )(t)  = \left\{ \begin{array}{lll} \gamma(2t) &{\rm if}& 0\le t<1/2\\
 \lambda(2t -1)& {\rm if}& 1/2 \le t \le 1~.
\end{array} \right. 
$$
The inverse $\gamma ^{-1}$ is defined by $\gamma ^{-1} (t)= \gamma (1 -t)$.

 By  immediate retracing we mean a parametrized  piecewise smooth path 
$\gamma$ which factorizes as
$$
\gamma =\prod _i ^k (\gamma _i \gamma _i ^{-1})~.
$$
 
Equivalence of piecewise smooth (or analytic) parametrized
  paths w.r.t. order preserving 
piecewise smooth  (or piecewise analytic) re\-pa\-ra\-me\-tri\-za\-tions and 
up to  immediate retracings will be called elementary equivalence.  A piecewise
 smooth (analytic) path  is an elementary equivalence class of  piecewise smooth 
(piecewise analytic) parametrized paths.   For applications to gauge theories it
 is convenient to consider just paths which are piecewise smoothly  immersed or 
constant. Paths can be 
composed  and inverted, and form a   groupoid, denoted by ${\sf Path}(M)$ in the
 piecewise analytic case  and  by $Path(M)$ in the piecewise smooth immersive
  case.  Obviously, one could also consider weaker differentiability conditions 
on paths and more general groupoids, as in \cite{Hyph}  and \cite{comm}.
 
 Fix a base point $\star$ in $M$.  A piecewise smooth (analytic)  loop 
 based on $\star~$ is an elementary equivalence class of piecewise smooth 
(analytic) parametrized paths $\gamma$ with $\gamma (0)=\gamma (1)=\star$.
 Loops based on
 $\star$ form a group, denoted  by ${\sf Loop}_\star (M)$ in the analytic case
 and  by  $Loop_\star (M)$ in the  smooth immersive case.  The unit is the 
 constant loop, also denoted by $\star$. If the base points are changed, one
 obtains  isomorphic groups.

\vskip 2 mm
  
 Let us consider now the trivial bundle $M\times G$, where
 $G$ is a  connected closed subgroup of $U({\sf n})$. For a smooth connection
 $A$ on $M\times G$,  the holonomy map $H_A $  is defined, which 
associates to a path $\lambda$ the parallel transport along $\lambda$, identified
 with an element $H_A (\lambda )$  of $G$. We  can identify the set  
${\mathcal A}$ of smooth connections  with the set of  their associated holonomy 
maps  on  ${\sf Path}(M)$, on $Path (M)$ \cite{giles} or even on more general
 groupoids. 

 Connections induce another equivalence relation on parametrized paths 
(of any class), called {\em holonomy equivalence}, where parametrized paths
 $\lambda$ and $\lambda^\prime$ are holonomy equivalent if 
 $$
\lambda ^\prime(0)= \lambda (0)~,~
\lambda ^\prime (1)= \lambda  (1) \quad {\rm and} \quad 
H_A (\lambda ^\prime \lambda^{-1}) = e \quad \forall A\in {\mathcal A}~.
$$ 
The holonomy equivalence depends, in principle, on $G$  and is weaker than 
  elementary equivalence. For every connected non-solvable compact (hence 
 non Abelian)  Lie group $G$,   the holonomy  equivalence agrees with elementary
 equivalence in the analytic and in the smooth immersive case 
 \cite{LewThiemann},\cite{Spallanzani}. In the smooth non-immersive case this
 is no longer true: a simple example  of a smooth loop giving a non trivial
 elementary equivalence class, but with trivial holonomies for any $G$, has 
 been given in  \cite{Hyph}.

For $G= T^n$, a torus in dimension $n,\quad n\ge 1$, the two equivalence relations
 are different. However, the holonomy equivalence does not depend on $n$. In the
 Abelian case, the group $Loop_\star (M)$ quotients to a group, we will denote by 
$Hoop _\star (M)$, since holonomy equivalence classes of loops were 
 first introduced in  \cite{AshtekarLewandowski} and  called hoops. 
Holonomy equivalence classes of paths in $Path (M)$ form a groupoid, which we
 analogously denote by $Hath (M)$.

\vskip 2mm
  The group of gauge transformations is the group 
$Gau = C^\infty (M, G)$, acting on $\mathcal{A}$ by $A.g=g^{-1}Ag +g^{-1}dg$ 
where $d$ denotes the exterior derivative. The corresponding action on parallel 
transports is given by
$$
H_{A.g} (\gamma )= g(\gamma (1))^{-1} H_A (\gamma ) g(\gamma (0)) \quad 
  \gamma \in Path(M)
$$
 and is the restriction to $Gau $ of the action of $G^M$. Of course,
 ${\mathcal A}$ is not $G^M$-invariant, however its $Gau$-invariance implies that 
${\mathcal A}_\star$ is $Ad G$-invariant. Since $Gau$ is dense in $G^M$ and the
 action of $G^M$ on $Hom(Path(M),G)$ is continuous, the $Gau$-invariance of 
${\mathcal A}$ implies the $G^M$-invariance of $\overline{\mathcal A}$.
 We recall that every function $f$ in $Cyl({\mathcal A})$ can be uniquely 
extended to a continuous function on $\overline{\mathcal A}$, which 
is $G^M$-invariant if $f$ is $Gau$-invariant. As a consequence, 
$Cyl_{Gau}({\mathcal A})= Cyl _{G^M} ({\mathcal A})$, in the notations of Theorem
 \ref{vattelapesca}.  

 The projection 
${\mathcal P}_\star :Hom (Path  (M) , G) \to Hom (Loop_\star (M) , G)$
 quotients to a bijection 
$$
{\mathcal A} /Gau \leftrightarrow {\mathcal A}_\star /Ad G~
$$
 (see \cite{Anandan}).
Analogous statements can be done using ${\sf Path}(M)$ and
 ${\sf Loop}_\star (M)$ and also for the weaker differentiability conditions
  on paths. 

\vskip 2mm

 In the non-perturbative quantization program  a standing point is to give a
 compactification of the configuration space ${\mathcal A}/Gau$, i.e. to  embed
 the configuration space in the spectrum of some $Ad G$-invariant cylindrical
 algebra on ${\mathcal A}_\star $. A compactification of ${\mathcal A}$  can be 
achieved by the embedding in the spectrum of the cylindrical algebra of 
 ${\mathcal A}$. One is interested in studying the consistence of these
 compactifications.

\vskip 2 mm
 
{\em The analytic case.} 

 In the analytic  case ${\mathcal A}$ is viewed as a subset 
of  $Hom ({\sf Path}(M), G)$. To distinguish this case from the smooth one,  we
 will denote by ${\sf Cyl}({\mathcal A})$ the analytic cylindrical algebra
 and by ${\sf Hol}({\mathcal A}_\star )$ the analytic holonomy algebra, i.e. 
 the   Wilson $\complessi^*$-algebra of ${\mathcal A}_\star$.

 In this setting the following strong version of the Approximation Condition
  on ${\mathcal A}$ is assured:
\par

\vskip 2 mm 
{\em  For every finite set of paths $\{ \gamma _k \}_{k=1,...,r}$  and every
 homomorphism $H: {\sf Path}(M) \to G$ there exists a  smooth connection $A$ on
 $M\times G$ such that
$$
H (\gamma _k)= H_A (\gamma _k) \quad k=1,...,r~.
$$}

The property has been proved  in \cite{AshtekarLewandowski}  for 
$\gamma_k\in{\sf Loop}_\star (M)$ ( see also \cite{Spallanzani}) and the proof
 extends easily to   $\gamma_k\in {\sf Path}(M)$.  Owing to Theorem 
\ref{cilindriche} we can conclude that  
$$
 Spec({\sf Cyl} ({\mathcal A}))\equiv \overline{\mathcal A}= 
 Hom ({\sf Path}(M),G)~.
$$

 As a consequence,  we obtain that
$$
Spec({\sf Cyl} _{Gau} ({\mathcal A})) \equiv 
Hom( {\sf Loop}_\star (M), G)/ AdG \equiv \overline{{\mathcal A}/ Gau}~.
$$

In the special case that $G$ is a closed normal subgroup of $U_{\sf n}$, the
   unitary equivalence classes of $G$-valued homomorphisms agree with their
  conjugation classes, so that 
$\mathcal{A} /Gau \equiv \widehat{{\mathcal A} _\star} =
 {\mathcal A}_\star / Ad G$. By Theorem \ref{stati} we get
$$
Spec({\sf Hol(\mathcal A}_\star))\equiv \widehat{Hom }({\sf Loop}_\star (M),G) =
\overline{{\mathcal A}/Gau}~.
$$
 Dually, we have that
$$
{\sf Hol}({{\mathcal A}_\star})={\sf  Cyl} _{Ad G} ({\mathcal A} _\star )
\equiv {\sf Cyl} _{Gau} ({\mathcal A})~.
$$
 
However, the analytic case is  not really satisfactory from the physical
 point of view, since for applications to the loop quantum gravity  one needs
to consider  diffeomorphism invariance  on the closure of ${\mathcal A}/Gau$ while 
${\sf Loop}_\star (M)$ is invariant only w.r.t. analytic diffeomorphisms.

\vskip 2 mm
{\em The  Abelian smooth immersive case.} 

Ashtekar and Lewandowski studied in \cite{AshtekarLewandowski} the case of 
$G=U(1)$, working in the piecewise $C^1$ setting. They denoted by
 $\mathcal H \mathcal G$ the group of hoops obtained by
 the  holonomy equivalence classes of piecewise $C^1$ loops based on $\star$  
 and considered in  the  compact space $Hom({\mathcal H}\mathcal G ,U(1))$ the
  subset ${\mathcal A}_\star$ obtained by the holonomy maps of the smooth
 connections. They proved that
$$
\overline{{\mathcal A}_\star }=Hom({\mathcal H{\mathcal G}},U(1))~.
$$
Their proof works also in the case of a general torus $T^n$ in any dimension $n$
 and for every differentiability class of loops and hoops. It works also 
 in the smooth immersive category, when the group $\mathcal H \mathcal G$ is
 replaced by the group $Hoop_\star  (M)$, so that
 $$
\overline{\mathcal A_\star} =Hom(Hoop_\star  (M),T^n).
$$
 Let us  consider the smooth connections $\mathcal A$ as  a subset of
 $Hom(Hath (M),T^n)$. The gauge invariance of 
${\mathcal A}$ implies the $G^M$-invariance of
 $\overline{\mathcal A}$. Moreover, ${\mathcal A}_\star$ is $AdG$-invariant.
   By the last statement in  Theorem \ref{vattelapesca} we obtain that 
$$
\overline {\mathcal A}=Hom(Hath (M),T^n) 
$$
and by  ${\mathcal A}/Gau \equiv {\mathcal A}_\star$ we have
$$
\overline {{\mathcal A}/ Gau } \equiv Hom(Hoop_\star (M),T^n)~. 
$$

The above quoted result allows one to characterize also the closure of
 $\mathcal A$ in the space $Hom(Path (M),T^n)$: we can  identify 
 $Hom(Hath (M),T^n)$ with the closed subgroup of 
 $Hom(Path (M),T^n)$, consisting of the homomorphisms which are constant on each 
holonomy equivalence class of paths, obtaining
$$
\overline{\mathcal A}=Hom(Hath (M),T^n)\subset Hom(Path (M),T^n)~.
$$

 Fleischhack  proves in \cite{tesi}  that $\overline{\mathcal A}$ is a proper
 subset in $Hom(Path (M),T^n)$. Analogously, one recognizes that
 $Hom (Hoop _\star (M) , T^n)$ is a closed proper 
 subgroup of $Hom (Loop _\star (M), T^n)$ and agrees with the closure  of  
${\mathcal A}_\star $ in $ Hom(Loop _\star (M),T^n)$. 
  
 In contrast, in the analytic case it is well 
known \cite{Spallanzani} that the group ${\sf Hoop}_\star (M)$ is the quotient of
 ${\sf Loop}_\star (M)$ by its commutator group, so that $
Hom ({\sf Loop}_\star (M), T^n)\equiv Hom ({\sf Hoop}_\star (M), T^n)
$ and, analogously, $Hom ({\sf Path} (M), T^n)\equiv Hom ({\sf Hath}(M), T^n)$.

\vskip 2 mm 

{\em The general  smooth immersive case.}

  Our aim is to characterize the closure of
 ${\mathcal A}$ in the smooth immersive case for a connected compact Lie 
 group $G$.  Fleischhack has proved that  ${\mathcal A}$ is dense in 
 $Hom (Path (M) , G )$ only for connected and  semisimple $G$ \cite{tesi}.
 The same author will discuss denseness of connections in the 
  non-immersive case and for more general categories of
 immersive paths in \cite{Fleish}.

The space $\overline {\mathcal A}$ was  investigated in the smooth immersive case
  for  any  connected compact Lie group $G$ by Baez and Sawin in \cite{BaezSawin1}.
They  associate to every finite ordered family $C=(c_1,...,c_r)$
of piecewise immersed or constant paths the map $p_{C}:{\mathcal A}\to G^r$,
 $H_C(A)=(H_{c_1}(A),...,H_{c_r}(A))$ and  studied its range ${\mathcal A}_C$.
 They found  special families of independent paths, the  webs, for which this
 range can be  characterized. We recall that a family $C$ of paths is said to be
 independent if a path $\lambda$ in  $ C$ cannot  be decomposed using other 
paths in $C$ or their inverses.  A family $C$ depends on another family $C^\prime$
 if  every path  in $C$ can be obtained from paths in $C^\prime$ or
 their inverses by using the path composition.

The definition of a web is quite involved and we refer to the quoted authors.
Their main result is that   every finite family $C$ of paths depends on a web.
 Moreover it was proved in \cite{LewThiemann} (see also \cite{Spallanzani}) that
  if $G$ is semisimple, for every web $W$ the range ${\mathcal A}_W$ is exactly
 $G^r$, where $r$ is the cardinality of $W$.

\vskip 2mm
Joining the above  results one obtains that 
$$\overline{\mathcal A}=Hom(Path(M),G)$$
 for every  connected compact semisimple Lie group $G$. 
\vskip 3mm

 Let now $G$ be isomorphic to the product of a torus $T^n$ and a connected compact 
 semisimple Lie group $S$. Then one obtains that
$$
Hom(Path(M),G) = Hom(Path(M),T^n )\times Hom(Path(M),S)
$$
 and that 
$$
Hom(Loop_\star (M),G)/Ad G= 
Hom(Loop_\star (M),T^n )\times Hom(Loop_\star (M),S)/Ad S~.
$$

 The Lie algebra $\mathfrak g$ splits as 
${\mathfrak g}={\bf R}^n +{\mathfrak s}$, so that a connection $A$ on
 $M\times G$ can be identified  with a couple $A_{T^n}$ and $A_S$ of 1-forms on 
$M$ taking values in  the Lie algebras ${\mathbf R}^n$ and $\mathfrak s$ of 
$T^n$ and $S$, respectively. It follows from the definition of path ordered
 integral that 
$
H_A(\lambda)=H_{A_{T^n}}(\lambda)H_{A_S}(\lambda)$ for every $ \lambda\in Path(M)
$.

\vskip 2mm

\begin{theorem}\label{riduttivo}
Let $G=T^n\times S$ the product of a torus $T^n$  and a connected compact
 semisimple Lie group $S$. The closure of the set $\mathcal A$ in  $Hom(Path(M),G)$
 is $ Hom(Hath (M), T^n )\times Hom(Path(M),S)$. The closure
of  ${\mathcal A}/Gau$ in  $Hom(Loop_\star (M),G)/Ad G$
 is $ Hom(Hoop _\star (M), T^n )\times Hom(Loop_\star(M),S)/Ad S$.
\end{theorem}
{\em Proof.} This is an immediate consequence of
 the above remarks and of  the results in the Abelian  and the semisimple cases, 
 respectively.  \hspace{\stretch{1}}$\Box$

\vskip 2 mm 
Let us recall that every compact connected Lie group $G$ is of the form 
$(T_0\times S)/K$ where $T_0$ is the identity component of the centre of $G$, $S$
 is a connected compact semisimple Lie group and $K$ is a finite group contained
 in the centre of $T_0\times S$. By the general theory of compact Abelian Lie
 groups, $T_0$ is trivial or it is a torus. We denote by $p_K$ the projection 
 $T_0 \times S \to G$ and by 
 $(p_K)_* : Hom (Path (M),T_0 \times S) \to Hom (Path (M), G)$ 
 the map defined by  $(p_K)_* (H)= p_K\circ H$.
  
The case where $T_0$ is a torus is discussed in the following theorem.

\vskip 2 mm 
\begin{theorem}\label{ultimo} Let $G$ be a compact connected Lie group represented 
 in the form $G= (T ^n\times S)/K$, as above.  The closure of $\mathcal A$   in 
 $Hom(Path(M),G)$ is $(p_K)_* (Hom( Hath (M), T^n )\times Hom (Path(M), S))$.
 The closure of ${\mathcal A}/Gau$ in $Hom ( Loop _\star (M), G)/Ad G$ is
  $(p_K)_* (Hom( Hoop_\star (M), T^n)\times Hom (Loop_\star (M), S))/Ad S)$.
\end{theorem}
{\em Proof.} The commutative diagram holds:
\begin{displaymath}
\begin{array}{ccc}
 {\mathcal B}     & {\longrightarrow}   & Hom(Path(M), T^n \times S) \\
\quad \quad\Big\downarrow {(p_K)_*}   &        &\quad\Big\downarrow{(p_K)_*}    \\
{\mathcal A}        & {\longrightarrow} & Hom(Path(M),G)
\end{array}
\end{displaymath}
where ${\mathcal B}$  denotes the   space of holonomies of smooth connections
 on $M\times (T^n \times S)$. Since the Lie algebras ${\bf R}^n + \mathfrak{s}$
 and $\mathfrak{g}$ are isomorphic, the bundles $M\times (T^n \times S)$ and
 $M\times G$ have isomorphic connection one-forms, hence
 $(p_K)_* ({\mathcal B})= {\mathcal A}$.
 
The map $(p_K)_*$ is continuous in the Tychonoff topologies and is  closed, since
  $Hom(Path(M),T^n \times S)$ is compact. Therefore 
$(p_K)_* (\overline{{\mathcal B}})= \overline{(p_K)_* ({\mathcal B})}= 
 \overline{\mathcal A}$. Then we use Theorem \ref{riduttivo}. 
 
To prove the second statement, we have to apply analogous arguments to
 the  diagram

\begin{displaymath}
\begin{array}{ccc}
 {\mathcal B}_\star /Ad S     & {\longrightarrow}   &
 Hom(Loop_\star (M), T^n \times S)/Ad S \\
\quad \quad\Big\downarrow {(p_K)_*}   &        &\quad\Big\downarrow{(p_K)_*}    \\
{\mathcal A}_\star /Ad G        & {\longrightarrow} & Hom(Loop_\star (M),G)/Ad G
 ~. \end{array}
\end{displaymath}
 
  The only non trivial point to prove  is  that the (quotiented) projection 
$(p_K )_*: {\mathcal B}_\star /Ad S \to {\mathcal A}_\star /Ad G$ is onto.  
 This follows immediately by $(p_K)_* ({\mathcal B}_\star )= {\mathcal A}_\star$. 
  \hspace{\stretch{1}}$\Box$

\vskip 2mm
We can  characterize the spectra of the smooth immersive cylindrical algebras 
 $Cyl ({\mathcal A})$ and $Cyl_{ Gau} ({\mathcal A})$ and
 of the smooth immersive holonomy algebra $Hol ({\mathcal A}_\star)$.  

 For a semisimple connected compact Lie group $G$ we have 
$$
 Spec(Cyl({\mathcal A}))\equiv \overline{\mathcal A}= Hom (Path (M) , G )~
$$
and
$$
 Spec(Cyl_{Gau}({\mathcal A}))\equiv\overline{{\mathcal A}/Gau}=
 Hom (Loop _\star (M) , G )/AdG~.
$$

 In the special case that  $G$ is a normal subgroup of $U({\sf n})$,
 e.g. if $G$ is $SU({\sf n})$, ${\sf n}>1$,  we identify 
$Hol({{\mathcal A}_\star})$ with $Cyl_{Gau}({\mathcal A})$ and we have
$$
Spec(Hol({{\mathcal A}_\star}))\equiv\overline{{\mathcal A}/Gau}=
\widehat{Hom}(Loop_\star(M),G)~.
$$

For $G= T^n \times S$, with semisimple $S$, we have  
 $$
 Spec(Cyl({\mathcal A}))=
 Hom (Hath (M) , T^n)\times Hom (Path (M), S)~
$$
and
$$
 Spec(Cyl_{Gau}({\mathcal A}))=
 Hom (Hoop _\star (M) , T^n ) \times Hom (Loop_\star (M), S)/AdS~.
$$

 Finally, for $G= (T^n \times S)/K$, as above, we have
$$
 Spec(Cyl({\mathcal A})) =
 (p_K )_*\left( Hom (Hath (M) , T^n)\times Hom (Path (M), S)\right) ~ 
$$
and
$$
 Spec(Cyl_{Gau}({\mathcal A}))=
 (p_K)_* \left(
 Hom (Hoop _\star (M) , T^n ) \times Hom (Loop_\star (M), S)/AdS
\right) ~.
$$

\section{Non  trivial bundles}

We will extend the above results to the  general  case that $\mathcal A$ is the 
set of smooth connections of a non trivial principal bundle $P(M,G)$. 

For $x,y\in M$ we denote by $Eq(P_x,P_y)$  the space of the $G$-equivariant maps
 from the fiber $P_x$ to the fiber $P_y$. This space consists of invertible maps
and  every choice of  $u_x\in P_x$ and $u_y\in P_y$ defines a  bijection of 
$Eq(P_x,P_y)$ onto $G$, so   $Eq(P_x,P_y)$ becomes a compact space whose topology
 does not depend on the choice. The disjoint union 
$ Eq(P)=\coprod_{x,y\in M}Eq(P_x,P_y)$ is a groupoid, since
 $\phi_{y,x}\in Eq(P_x,P_y)$ and  $\phi_{y',x'}\in Eq(P_{x'},P_{y'})$ can be 
composed if $y=x'$.

 Let we denote by $\Lambda $ any of the groupoids of paths previously introduced
 and consider the space  $Hom(\Lambda ,Eq(P))$ of groupoid homomorphisms  of 
 $\Lambda$ in $Eq(P)$. The space $ Hom(\Lambda ,Eq(P))$ is compact as a closed 
subset of the product 
$\prod _{\lambda \in \Lambda} Eq(P_{s(\lambda )},P_{r(\lambda )})$, which
 is compact in the Tychonoff topology.

 We fix a point $\star$ in $M$ and introduce  the compact space 
${\mathcal E}_\star$ of the maps $\eta \in \prod _{x\in M} Eq(P_\star,P_x)$ 
satisfying  $\eta(\star)=id$.  For every $x\in M$,
 we fix a path  $e_x\in \Lambda $ with $s(e_x)=\star$ and  $r(e_x)=x$, choosing
 $e_\star=\star$. To every   $H\in Hom(\Lambda ,Eq(P))$  we associate the element
$\eta_H\in{\mathcal E}_\star $ given by $\eta_H(x)=H(e_x)$.

 Fixed a point $u_\star$ in $P_\star$,  we identify the group
 $Eq(P_\star,P_\star)$ with the group $G$ and define  the map 
 ${\mathcal P}_\star:Hom(\Lambda ,Eq(P))\to Hom(\Lambda _\star ,G)$ 
which associates to each $H\in Hom(\Lambda ,Eq(P))$ its restriction to 
$\Lambda _\star $.

 As in section 2, the map $ H\mapsto ( {\mathcal P}_\star(H), \eta_H )$ is a
 homeomorphism
$$
\Theta  : Hom(\Lambda  , Eq(P)) \to Hom (\Lambda _\star , G)
 \times {\mathcal E}_\star ~.
$$ 
 
Let us recall that $Gau (P)$ is the group of the smooth sections of the fiber
 bundle $\coprod_{x \in M} Eq(P_x , P_x ) \to M$, with fiber $G$.  $Gau(P)$ is 
 dense in the group $\overline{Gau (P)}$ of all sections of this bundle, a
 compact group isomorphic to  $G^M$  (see \cite{ALpro}).

Actions of $\overline{Gau (P)}$  on  $Hom(\Lambda ,Eq(P))$, on
 $ Hom (\Lambda _\star , G)$ and on ${\mathcal E}_\star$ are defined as follows:
$$\begin{array}{ll}
(H\phi)(\lambda)=\phi(r(\lambda))^{-1}H(\lambda)\phi(s(\lambda))& H\in
Hom(\Lambda ,Eq(P)),\\

 (H_\star\phi)(\gamma)=Ad_{\phi(\star)^{-1}}H_\star(\gamma)& H_\star\in 
Hom (\Lambda _\star , G),\\

 (\eta\phi)(x)=\phi^{-1}(x)\eta(x)\phi(\star) & \eta\in {\mathcal E}_\star ,
\end{array}
$$
for every $\phi\in \overline{Gau (P)}$.

The map $\Theta$ is equivariant and quotients to a homeomorphism of 
$ Hom(\Lambda , Eq(P))/\overline{Gau(P)}$ onto  $Hom (\Lambda _\star , G)/AdG$.

 Cylindrical functions and algebras can be treated  for the space $\mathcal A$ of
 connections on a general bundle.  A cylindrical function is a function of the form
$F_{\lambda_1, ...,\lambda_r ;f}(A)=f(H_A(\lambda_1) ,...,H_A(\lambda_r))$, where
 $f$ is a continuous function on the compact space 
$Eq(P_{s(\lambda _1)},P_{r(\lambda _1)})\times...\times 
Eq(P_{s(\lambda _r)},P_{r(\lambda _r) })$.

 Analogous embedding results to the ones given in section 2 and section 3 can be easily 
worked out for the spectrum of these cylindrical algebras.  Analogously to the trivial case, 
 we obtain  
$$
Spec(Cyl({\mathcal A}))\equiv \overline{\mathcal A} \subset Hom (\Lambda, Eq (P))
~,
$$
and
$$Spec (Cyl_{Gau(P)}({\mathcal A}))\equiv \overline{{\mathcal A}/Gau (P)} 
\subset Hom (\Lambda _\star , Eq (P))/Ad G~.
$$

For the groups $U({\sf n})$ or $SU({\sf n})$ we have that 
$$
Hol({\mathcal A_\star})=Cyl_{Gau(P)}({\mathcal A})
$$
and that
$$
  Spec(Hol({\mathcal A_\star})) \equiv \overline{{\mathcal A}/Gau(P)}  
 \subset Hom (\Lambda _\star , Eq (P))/Ad G~~.
$$

 However, the concrete characterization of  the closure of ${\mathcal A}$ in the non trivial case
 remains an open problem.  In the analytic case, the Approximation Condition on
 ${\mathcal A}$ is proved for any $G$ (see e.g. \cite{Spallanzani}).  
 In the smooth immersive case, the Approximation Condition can be proved
  for a semisimple group $G$ using the fact that webs are decomposed in tassels,
 each  contained in some trivializing neighborhood, as  in the proof of
 Proposition 2 in \cite{BaezSawin1}. 
 
 The results obtained in the Abelian non-analytic case are difficult to be
  generalized to non trivial bundles, since hoops are  not local. 
  Ashtekar and Lewandowski proved that 
$\overline{{\mathcal A} _\star} = Hom ({\mathcal H}{\mathcal G}, U(1))$  for 
  the Hopf bundle and its pullbacks \cite{AshtekarLewandowski}. 

\vskip 1 cm
{\bf Acknowledgements.} The authors would like to thank C. Fleisch\-hack for his  
 comments on the manuscript and for signaling out his results contained 
in \cite{tesi}, \cite{Hyph} and \cite{Fleish}.



\begin{thebibliography}{99}


\bibitem{AbbatiMania} M. C. Abbati and A. Mani\`a, { On differential 
structure for projective limits of manifolds}, J. Geom. Phys. 
{\bf 29} (1999), 35-63.

\bibitem{AMP} M.C.  Abbati, A. Mani\`a and E. Provenzi, {Inductive 
Construction of the Loop Transform for Abelian Gauge Theories,} Lett. Math. Phys.
 {\bf 57} (2001), 69-81.


\bibitem{Anandan} J. Anandan, { Quantum Interference and the Classical 
Limit}, Int. J. Theor. Phys. {\bf 19} N. 7 (1980), 537-556.



\bibitem{AshtekarIsham} A. Ashtekar and  C. Isham,  Representations of the 
holonomy algebras of gravity and non-Abelian gauge theories, Class. 
Quant. Grav. {\bf 9} (1992), 1433-1485.

\bibitem{Ashtekar} A. Ashtekar, J. Lewandowski, D. Marolf, J.Mour\~{a}o
 and T. Thiemann, { Quantization of diffeomorphism invariant theories of
 connections with local degrees of freedom}, J. Math. Phys. {\bf 36} 11 
(1995), 6456-6493.  

\bibitem{AshtekarLewandowski} A. Ashtekar and J. Lewandowski, { 
Representation Theory of analytic holonomy $\complessi ^*$-algebras},
 in: J. C. Baez (ed.), {\em  Knots and Quantum Gravity},  Oxford Univ. Press. (1994) pp. 21-61.

\bibitem{ALpro} A. Ashtekar and J. Lewandowski, { Projective techniques and functional integration for gauge theories},  J. Math. Phys. {\bf 36} (1995), 2170-2191.


\bibitem{Baez1} J. C. Baez,  Diffeomorphism-invariant Generalized Measures
 on the Space of Connections Modulo Gauge Trasformations, in L. Crane and D. Yetter (eds.),  {\em The Proceedings 
of the Conference on Quantum Topology}, World Scientific (1994), pp. 213-223.


\bibitem{BaezSpin}J. C. Baez, Spin Networks in Nonperturbative Quantum Gravity, in L. Kaufmann (ed.), {\em The interface of Knots and Physics},  A.M.S. (1996), pp. 167-203.

\bibitem{BaezSawin1} J. C. Baez and S. Sawin, { Functional Integration on
spaces of Connections}, J. Funct. Analysis {\bf 150} (1997), 1-26.

\bibitem{BaezSawin2} J. C. Baez and  S. Sawin, { Diffeomorphism-Invariant
Spin Network States}, J. Funct. Analysis {\bf 158} (1998) 253-266.


\bibitem{Baumgartel} H. Baumg\"artel, { On a theorem of Ashtekar and 
Lewandowski in the mathematical framework of canonical quantization of 
quantum gravity}, in: H. J. Schmidt and M. Rainer (eds),  {\em Current Topics in Mathematical Cosmology},  Proc. Int. 
Seminar, World Scientific
 (1998), pp. 217-222.


 
\bibitem{Broecker} T.Br\"ocker, T.tom Dieck,
 {\em Representations of Compact Lie Groups}, G.T.M. {\bf  98} Springer (1995).



\bibitem{Dixmier} J. Dixmier, {\em Les ${\mathbb{C}}^*$-alg\`ebres et leurs
 repr\'esentations},  Gautier-Villars (1969).



\bibitem{Eilenberg} S. Eilenberg and N Steenrod, {\em Foundations of Algebraic
 Topology}, Princeton University Press (1952).


\bibitem{tesi} C. Fleischhack,  {Mathematische und physikalische Aspekte
 verallgemeinerter Eichfeldtheorien im Ashtekarprogramm}, PhD thesis, Universit\"at
 Leipzig (2001). 

\bibitem{Hyph} C. Fleischhack,  { Hyphs and the Ashtekar-Lewandowski Measure}, 
to appear in J. Geom. and Phys. arXiv:math-ph/0001007. 

\bibitem{comm} C. Fleischhack, {Stratification of the Generalized Gauge 
Orbit Space}, Commun. Math. Phys. {\bf 214} (2000) 607-649.  



\bibitem{Fleish} C. Fleischhack, { Regular Connections among Generalized 
 Connections}, in preparation. 




\bibitem{giles} R. Giles, { Reconstruction of gauge potentials from Wilson
  loops,} Phys. Rew. D {\bf 24} N. 8 (1981),  2160-2168.  








\bibitem{Kobayashi} S. Kobayashi and K. Nomizu, {\em Foundations of 
Differential Geometry}, Vol I, Wiley (1963).


 

\bibitem{LewThiemann} J. Lewandowski and T. Thiemann, { Diffeomorphism
invariant Quantum Field Theories of Connections in terms of webs}, 
Class. Quant. Grav. {\bf 16} (1999), 2299-2322.









\bibitem{Misuranulla} D. Marolf and J. M. Mour\~ao, { On the Support of the 
Ashtekar-Lewandowski Measure},  Comm. Math. Phys. {\bf 170} (1995), 583-605.

 




\bibitem{Proprietamisura} J. M. Mour\~ {a}o, T. Thiemann and J. M. Velhinho, 
{ Physical Properties of Quantum Field Theory Measures}, J. Math. Phys.
 {\bf 40} (1999) 2337-2353.


\bibitem{Rovelli} C. Rovelli and L. Smolin, { Loop Space Representation of 
Quantum General Relativity}, Nuclear Physics B331 (1990), 80-152.


\bibitem{Spallanzani} P. Spallanzani, Groups of loops and hoops, Comm. Math. Phys. {\bf 216} (2001), 243-253.



\bibitem{Velhinho} J. M. Velhinho, { A groupoid approach to spaces of generalized
 connections}, arXiv:hept-th/0011200 .





\end{thebibliography}
\end{document}